\documentclass[12pt]{article}
\usepackage{fullpage,setspace}
\def\gsim{\:\raisebox{-0.5ex}{$\stackrel{\textstyle>}{\sim}$}\:}
\advance \voffset by -.3cm
\advance \hoffset by -.5cm
\begin{document}
\hrule width 0pt
\vspace{3.5cm}
\begin{center}
CONSTRAINTS ON THE CHARGED HIGGS SECTOR FROM THE CDF TOP QUARK
DATA \\
\bigskip

D.P. Roy \\ T.I.F.R., Homi Bhabha Road, Bombay 400 005, India
\end{center}

\vspace{5cm}

\begin{abstract}The top quark data in the lepton plus $\tau$ channel offers
a viable probe for the charged Higgs boson signal.  We have analysed
the recent Collider Detector at Fermilab (CDF) data in this channel to
obtain a significant limit on the $H^\pm$ mass in the large
$\tan\beta$ region.
\end{abstract}

\newpage

\onehalfspace

The discovery of the top quark at the Tevatron Collider has led to a
good deal of interest in the search for new particles in top quark
decay$^{1]}$.  In particular the top quark decay is known to be a
promising place to look for the charged Higgs boson$^{2]}$, occurring
in the Minimal Supersymmetric extension of the Standard Model.  The
MSSM contains two Higgs doublets of opposite hypercharge; and the
ratio of their vacuum expectation values defines the parameter
\begin{equation}
\tan\beta = \langle\phi^0_2\rangle / \langle\phi^0_1\rangle.
\end{equation}
The 2 complex doublets correspond to 8 independent states, 3 of which
are eaten up by the $W^\pm$ and $Z^0$ bosons.  This leaves 5 physical
states --- the 3 neutral scalars $H^0_1, H^0_2, A^0$ along with the
charged pair 
\begin{equation}
H^\pm = \cos\beta \phi^\pm_2 - \sin \beta\phi^\pm_1.
\end{equation}
In the diagonal KM matrix approximation, the $H^\pm$ couplings to the
important quark and leptonic channels are given by
\begin{equation}
{\cal L} = {g \over \sqrt 2 M_W} H^+\Big\{\cot\beta m_t \bar
t b_L + \tan\beta  m_b  \bar tb_R + \cot\beta m_c \bar c s_L
+ \tan \beta m_\tau \bar\nu \tau_R\Big\} + hc,
\end{equation}
from which one can calculate the branching ratios $B(t \rightarrow
bH^+)$ and $B(H^+ \rightarrow \tau \nu)$.  The leading log QCD
correction is taken into account by substituting the quark mass terms
in (3) by their running masses$^{3]}$
\begin{equation}
m_q = m_q (M_{H^\pm})
\end{equation}
Fig. 1 shows the resulting branching ratios $B(t \rightarrow bH^+)$
and $B(H^+ \rightarrow \tau^+\nu)$ against $\tan\beta$ as solid and
dashed lines respectively for a representative $H^\pm$ mass of 100
GeV.  The former is peaked at small ($\leq 1$) and large ($\geq
m_t/m_b$) values of $\tan \beta$, while the latter reaches its maximal
value of 1 for $\tan\beta \geq 2$.  It is the product of these two
branching ratios that determines the size of the effective $H^\pm$
signal.  Thus one expects a viable signal in the region of large
$\tan\beta(\geq 40)$.

\newpage

The top quark pair are produced via quark-antiquark or gluon-gluon
fusion and decay via the $W^\pm$ or $H^\pm$ bosons.  While $H
\rightarrow \tau\nu$ decay has a branching ratio $\sim 1$ for
$\tan\beta \geq 2$, the branching ratios for $W$ decays are 
\begin{equation}
W \rightarrow \ell\nu (2/9), \tau\nu (1/9), \bar qq(6/9).
\end{equation}
Requiring leptonic ($e$ or $\mu$) decay of one $W$ boson for a viable
signal leads to the following final states,
\begin{equation}
\bar t t \rightarrow \bar bb\pmatrix{W & W, && W & H,&&  HH \cr
\downarrow & \downarrow && \downarrow & \downarrow && \cr
\ell\nu & \bar qq & (24/81) & \ell\nu & \tau\nu & (4/9) & \cr
& \tau\nu & (4/81) &&&& \cr
& \ell\nu & (4/81) &&&& \cr}
\end{equation}

Thus for the SM decay via $W$, the $\bar tt$ signal occurs in the (I)
$\ell$ + multi-jet and (II) $\ell + \tau$ channels with relative
probabilities of 24/81 and 4/81 respectively.  However in the presence
of the $t \rightarrow bH$ decay, the latter channel has a sizeable
additional contribution with a relative probability of 4/9 times $B(t
\rightarrow bH)/B(t \rightarrow bW)$.  The $\ell$ + multi-jet channel
(I) is the most significant channel for the $\bar tt$ signal, where
the CDF experiment has seen $\sim$ 26 signal events using SVX
$b$-tagging$^{1]}$. Using this for normalisation, one can predict the
number of $\bar tt$ events in the $\ell + \tau$ channel (II) provided
one knows the relative detection efficiencies for the two channels. 

We have made a Monte Carlo estimate of the detection efficiencies for
the (I) $\ell$ + multi-jet channel with SVX $b$-tagging and (II) $\ell
+ \tau$ channel with hadronic decay of $\tau$, incorporating the CDF
cuts and identification efficiencies in each case$^{1,4]}$.  We
get$^{5]}$
\begin{equation}
{\cal E}_I(WW) = 0.12, {\cal E}_{II}(WW) = .037, {\cal E}_{II} (WH) =
.047.
\end{equation}
The last number corresponds to a $H^\pm$ mass of 100 GeV, but rather
insensitive to this parameter.  The larger efficiency for the $WH$
compared to the $WW$ contribution in the $\ell + \tau$ channel is
mainly due to the opposite polarization of $\tau$ in the two
cases$^{5,6]}$.

\newpage

Fig. 2 shows the predicted $\bar tt$ signal cross-section in the
$\ell+\tau$ channel against $\tan\beta$ for different $H^\pm$
masses$^{5]}$.  The right scale shows the corresponding number of
events for the CDF integrated luminosity of 110 $pb^{-1}$$^{1]}$.
While one expcects only $\sim 1.5$ events from the $WW$ contribution,
there can be a large contribution from $WH$ at $\tan\beta \gsim 40$.
The CDF experiment has reported 4 events against a non-top background
of $2 \pm .35$$^{1,4]}$.  the corresponding 95\% C.L. limit of 7.7
events is indicated by the dashed line.  It implies a lower limit of
$M_{H^\pm} > 100~{\rm GeV}$ for $\tan\beta > 40$. 

Fig. 3 shows the 95\% CL limit on $H^\pm$ mass against
$\tan\beta$$^{5]}$.  The right scale shows the corresponding limit on
the $A^0$ mass.  Thus the CDF data seems to rule out a light
pseudoscalar $(M_{A^0} < 60~{\rm GeV})$ for $\tan\beta \geq 40~{\rm
GeV}$, which was required for the so called large $\tan\beta$ solution
to the (so called!) $R_b$ anomaly$^{7]}$.

The above limit is significantly stronger than the one recently
obtained by the CDF collaboration on the basis of their inclusive
$\tau$-channel data$^{8]}$.  However, the difference can be largely
traced to the different normalisation method adopted there$^{8]}$.  They
use the SM cross-section for $\sigma(\bar tt)$ for normalisation,
instead of the experimental cross section for $\sigma(\bar tt
\rightarrow \bar bbWW)$, as measured via their $\ell$ + multijet
events.  We feel that the normalisation method based on the latter is
less model dependent and more powerful at large $\tan\beta$.

The work reported here was done in collaboration with Dr Monoranjan
Guchait. 

\bigskip

\singlespace
\noindent{\large References}

\begin{enumerate}
\item P. Tipton, 28th Intl. conf. on High Energy Physics, Warsaw (July
1996).
\item V. Barger et.al., Phys. Rev. {\bf D41}, 3241 (1990);
R.M. Godbole and D.P. Roy, Phys. Rev. {\bf D43}, 3640 (1991).
\item M. Drees and D.P. Roy, Phys. Lett. {\bf B269}, 155 (1991).
\item CDF Collaboration: M. Hohlman, Lake Louise Winter School (1996);
S. Leone, XI Workshop on $\bar pp$ Coll. Phys., Padova (1996).
\item M. Guchait and D.P. Roy, hep-ph/9610514, Phys. Rev. D (in
press).
\item S. Raychaudhuri and D.P. Roy, Phys. Rev. {\bf D52}, 1556 (1995)
and {\bf D53}, 4902 (1996).
\item D. Garcia, R. Jimenez and J. Sola, Phys. Lett. {\bf B347}, 321
(1995); P.H. Chankowski and S. Pokorski, Nucl. Phys. {\bf B475}, 3
(1996).
\item CDF Collaboration: C. Loomis, PDF meeting, Minneapolis (1996),
Fermilab Conf. 96/232-E.
\end{enumerate}
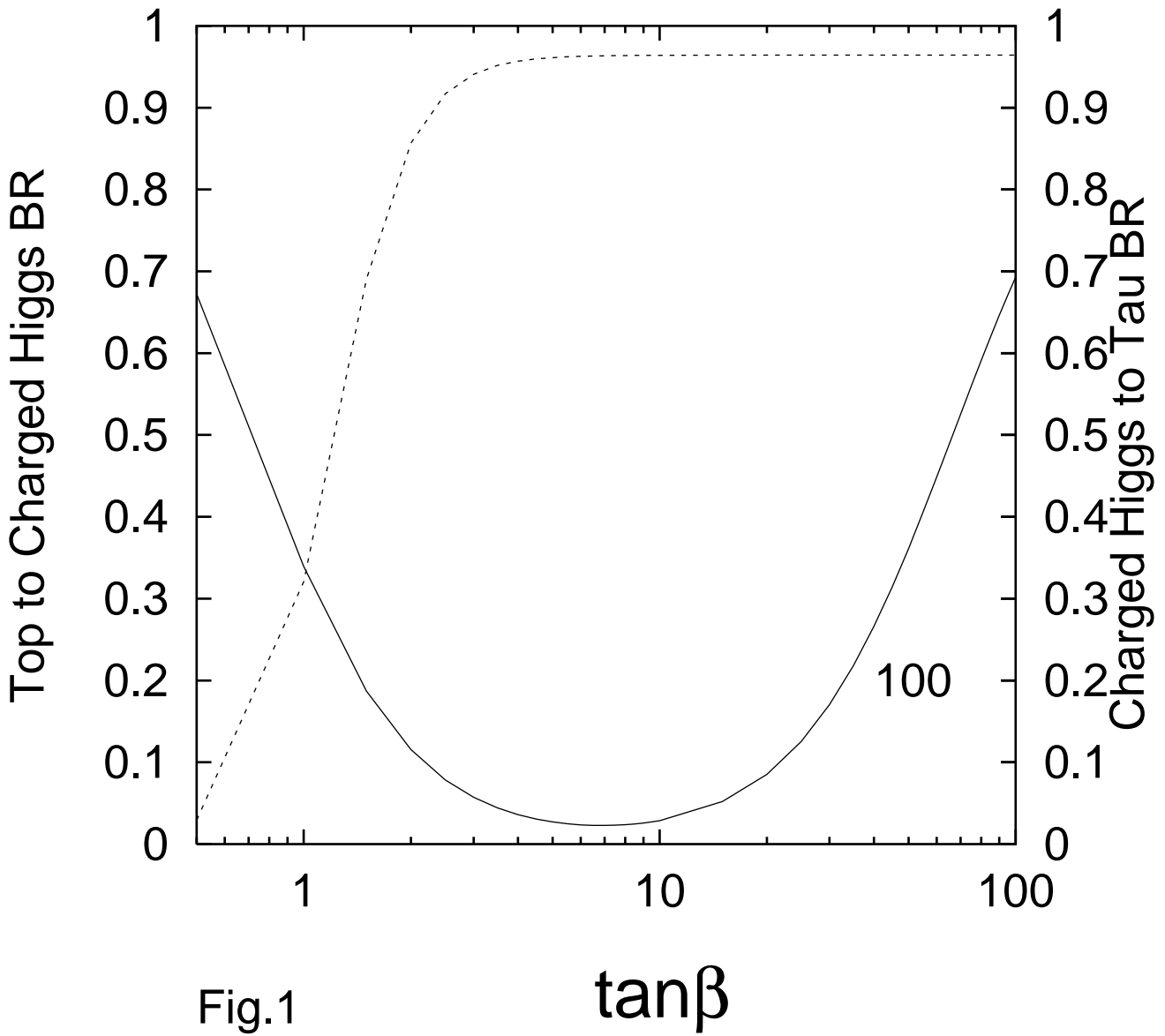
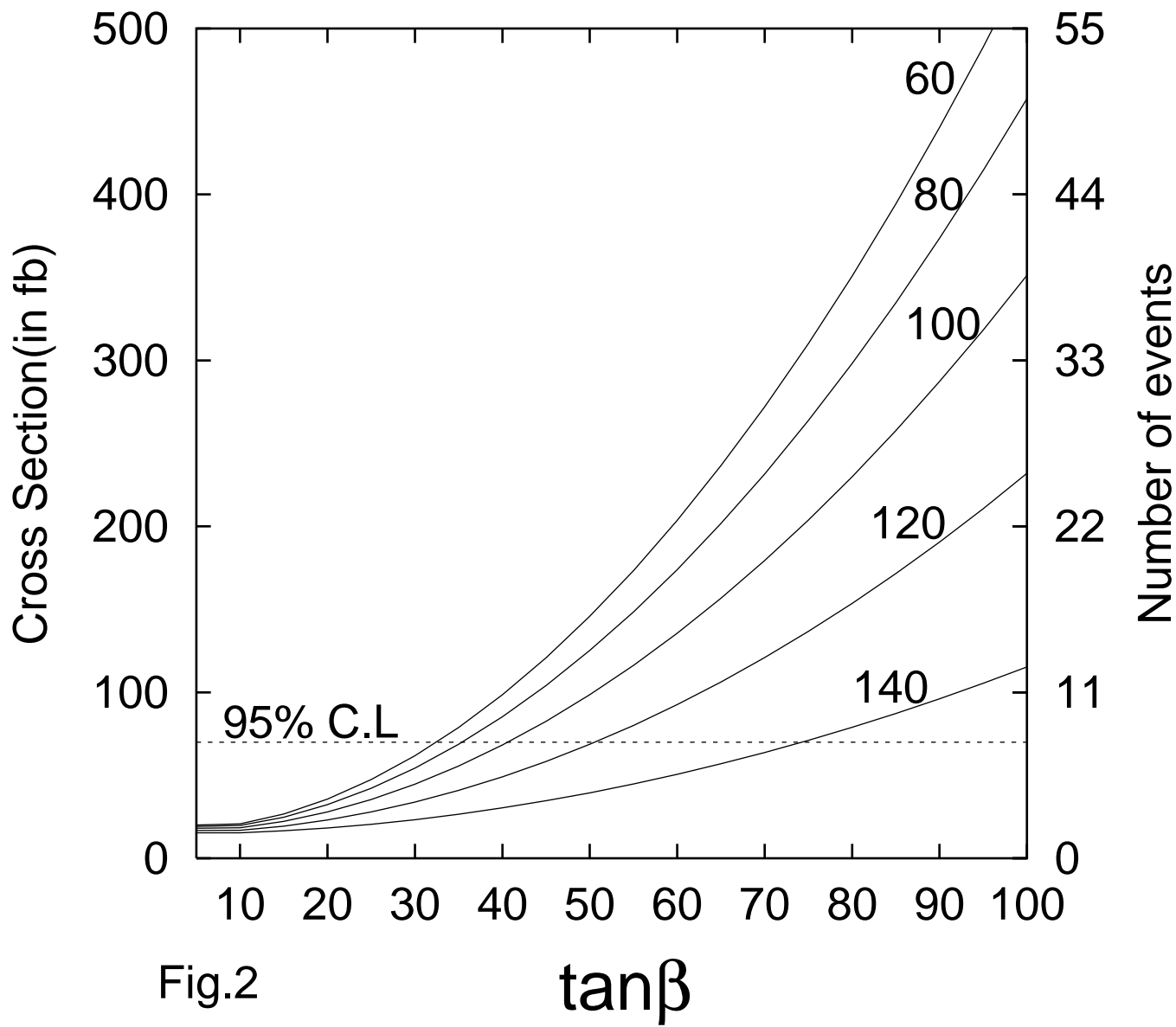
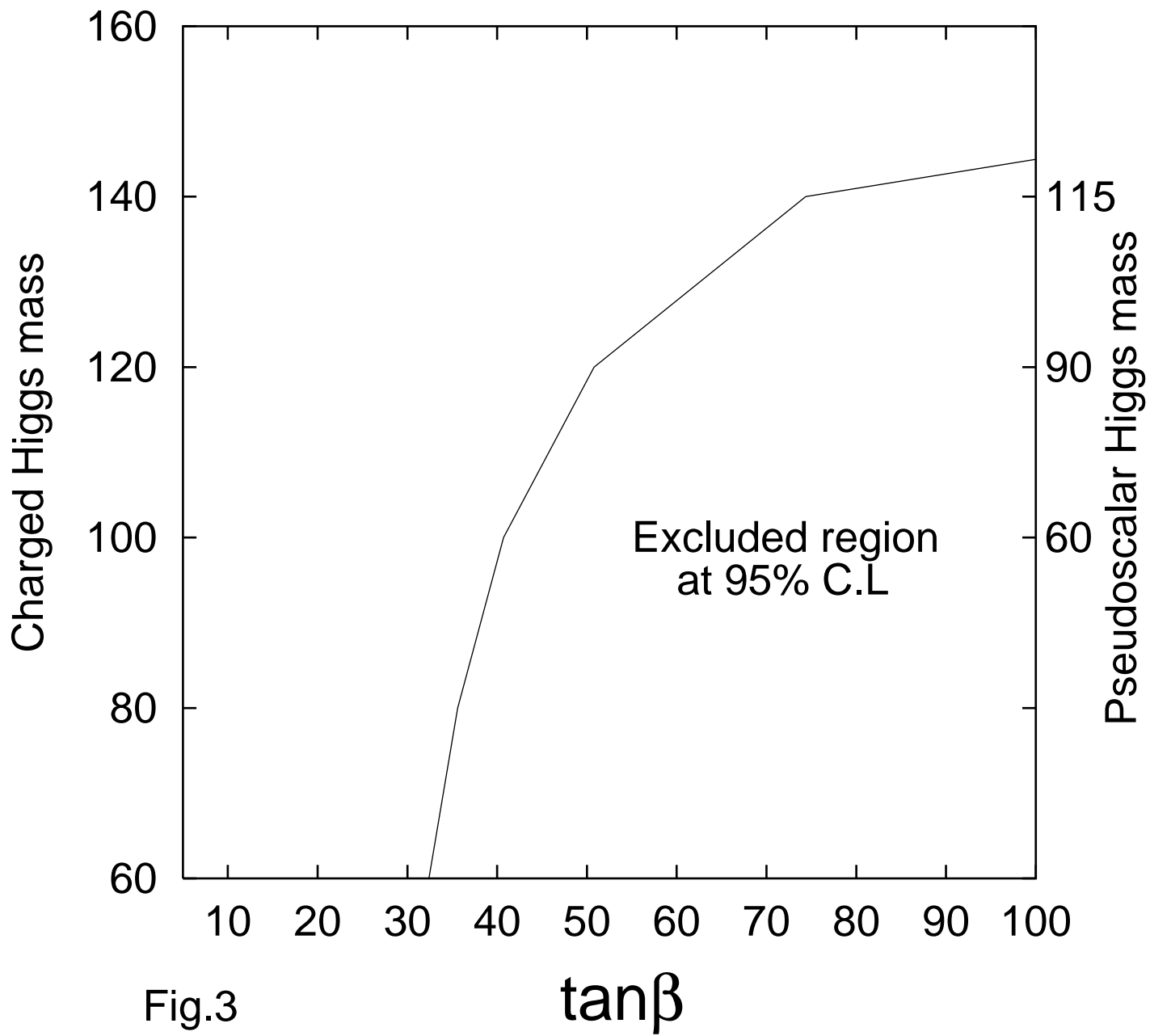
\end{document}